\documentclass[12pt]{iopart}
\usepackage{epsfig,graphicx,subfigure,amssymb,dcolumn,bm,multirow}

\begin{document}

\title{Hyperlink prediction via local random walks and Jensen-Shannon divergence}

\author{Xin-Jian Xu$^{1}$, Chong Deng$^{1}$ and Li-Jie Zhang$^{2}$}

\address{$^{1}$Department of Mathematics, Shanghai University, Shanghai 200444, People's Republic of China\\
$^{2}$Department of Physics, Shanghai University, Shanghai 200444, People's Republic of China}
\ead{lijzhang@shu.edu.cn}

\begin{abstract}
Many real-world systems involving higher-order interactions can be modeled by hypergraphs, where vertices represent the systemic units and hyperedges describe the interactions among them. In this paper, we focus on the problem of hyperlink prediction which aims at inferring missing hyperlinks based on observed hyperlinks. We propose three similarity indices for hyperlink prediction based on local random walks and Jensen-Shannon divergence. Numerical experiments show that the proposed indices outperform the state-of-the-art methods on a broad range of datasets.
\end{abstract}


\maketitle

\section{Introduction}

Graphs, or networks, are powerful tools for modeling complex systems, with vertices representing individuals and edges describing pair-wise interactions between two vertices. However, in systems such as co-authorship networks, metabolic networks and contact networks, interactions may involve more than two systemic units. Hypergraphs, or hypernetworks, which allow an edge (called hyperedge or hyperlink) to connect arbitrarily many vertices, are more suitable for modeling such systems than graphs~\cite{battiston2020networks}. For example, in a co-authorship network, an article co-authored by multiple authors can be conveniently represented by a hyperedge connecting all of its authors. Instead, connecting all pairs of authors by dyadic edges will cause information loss: we cannot tell from these dyadic edges which authors co-authored the article.

In the study of complex networks, one of the most important topics is link prediction~\cite{lv2011survey,wu2016}. A large number of studies have devoted to the literature with focus on pairwise interactions~\cite{kumar2020link,zhou2021review}. Until recently, little attention has been paid to hyperlink prediction, which is a natural extension of link prediction. Xu et al.~\cite{xu2013hyperlink} exploited the homophily property of social networks and developed a latent social feature learning scheme for hyperlink prediction. Zhang et al.~\cite{zhang2018beyond} formulated hyperlink prediction as a classification problem and adopted coordinated matrix minimization (CMM) to infer most likely missing hyperlinks from the candidate hyperlink set. Assuming that hyperlinks more often form from cliques than non-cliques, Sharma et al.~\cite{sharma2021c3mm} further improved the CMM with the introduction of \lq\lq clique-closure\rq\rq. On the other hand, Kumar et al.~\cite{kumar2020hpra} proposed a similarity-based hyperlink prediction method by considering resource allocation process in hypergraphs, called HPRA. The HPRA significantly outperforms those naive generalizations of similarity-based link prediction methods and is able to predict hyperlinks without any candidates. More recently, Pan et al.~\cite{pan2021predicting} proposed a topological-feature-based approach by extending loop-based link prediction method to hypernetworks, which is time-consuming when the size of the hypernetwork is large.

In view of the fact that naive generalizations of link prediction methods may exhibit poor performance in hyperlink prediction and promising performance can be achieved by utilizing the hypernetwork structure, we study random-walk-based link prediction in hypergraphs. It has been demonstrated that local random walks (LRWs) can achieve high prediction accuracy with low computational complexity in link prediction~\cite{liu2010link}. In this paper, we first consider local random walks on hypergraphs and associate a probability distributions to each vertex in the hypergraph. Then we measure hyperedge scores by either the corresponding transition probability distributions or the the Jensen-Shannon divergence between theses distributions. We demonstrate the superior performance of our indices over several promising baselines.

\section{Hyperlink prediction}\label{preliminary}

An undirected unweighted hypergraph is a tuple $G=(V,E)$ where $V$ is the set of vertices (or nodes) and $E$ is the set of hyperedges (or hyperlinks). Each hyperedge $e$ is a subset of $V$ and we assume that each hyperedge contains at least two vertices. Let $n=|V|$ and $m=|E|$ respectively be the number of vertices and hyperedges in the hypergraph. Without loss of generality, we denote $V=\{1,2,\cdots,n\}$. The incidence matrix of the hypergraph is a $n\times m$ matrix $H$ where $h_{ie}=1$ if $i\in e$ and $h_{ie}=0$ otherwise. The degree of vertex $i$ is defined by $d_i=\sum_{e\in E}h_{ie}$ and the cardinality of hyperedge $e$ is $|e|=\sum_ih_{ie}$.

Let $D_v\in\mathbb{R}^{n\times n}$ and $D_e\in\mathbb{R}^{m\times m}$ be diagonal matrices with veterx degrees and edge cardinalities in the main diagonals, respectively. The adjacency matrix of $G$ is $A=HH^T-D_v$ where $a_{ij}$ is the number of hyperedges that contain both $i$ and $j$. The weighted adjacency matrix of $G$ is $W=H(D_e-I)^{-1}H^T-D$ where $I$ is the identity matrix and $D=\mathrm{diag}(H(D_e-I)^{-1}H^T)$. Both $A$ and $W$ are symmetric and the diagonal elements are zero. Moreover, the row sums of $W$ are exactly the vertex degrees.

For link prediction, one usually assigns a score to each potential edge to indicate how likely it is a missing link, and choose the edges with the highest scores as our prediction. However, since the number of possible hyperlinks of a hypergraph is exponential in the number of vertices, this approach is infeasible for hyperlink prediction. Noting that in practice many potential hyperedges are highly irrelevant and thus can be easily filtered out, the hyperlink prediction problem is generally posed with a candidate hyperedge set~\cite{zhang2018beyond}: given an incomplete hypergraph and a candidate hyperlink set, hyperlink prediction aims at finding the most likely missing hyperedges from the candidate hyperlinks. The candidate set is usually built by expertise knowledge or negative sampling~\cite{patil2020negative} which generate non-existent hyperlinks based on the observed ones. Similar to link prediction, hyperlink prediction methods also first assign a score to each candidate hyperlink, and then select the hyperlinks with the highest scores as the prediction.

Unlike the situation in graphs where the random walk process is uniquely determined by the adjacency matrix, there are multiple ways to define random walks on hypergraphs. In the present work we consider the random walk in its simplest form~\cite{mulas2022random}. (i) A random walker sitting at vertex $i$ first walks to an incident hyperedge $e$ with equal probability. (ii) The walker then walks to vertex $j\in e$ other than $i$ ($j\neq i$) with equal probability. The transition matrix $P$ of this random walk has entries
\begin{equation}
p_{ij}=\frac{1}{d_i}\sum_{e\in E} \frac{h_{ie}h_{je}}{|e|-1}\cdot\textbf{1}_{i\neq j},
\end{equation}
where $\textbf{1}$ is the indicator function. In matrix form, we have $P=D_v^{-1}W$. The above hypergraph random walk is equivalent to a random walk on an undirected weighted graph whose adjacency matrix is $W$, which is called weighted projection of the hypergraph. The random walk process defined above assigns lower transition probabilities to vertices that are part of large hyperedges. An opposite choice was made in Ref.~\cite{carletti2020random}, which assigns higher transition probabilities to vertices belonging to large hyperedges and thus defines another projection of the hypergraph. Choosing an appropriate random walk process is an important but understudied problem. Here, we choose the former one since vertex hyperdegrees, which we consider to be an important property of the hypergraph, remain unchanged after projection.

The basic idea behind predicting missing (hyper)links via random walks is that tightly connected vertices are more likely to form (hyper)links and random walkers are more likely to transit between those closely connected vertice. Notice that $(P^k)_{ij}$ is exactly the probability that a random walker starting from $i$ reaches $j$ after $k$ hops. Hence we can expect that a larger value of $(P^k)_{ij}$ indicates stronger connection between $i$ and $j$. In fact, this only holds when $k$ is small because $(P^k)_{ij}$ converges to $d_j/(\sum_xd_x)$ as $k$ becomes large. This motivates us to measure vertex similarity using $P^k$ for small $k$, which implies that only local random walks are considered.

The next question is how to combine the vertex similarity information contained in random walks of different lengths. Following the idea of Liu and L\"{u}~\cite{liu2010link}, we treat random walks of different lengths equally and superpose their contributions, and define
\begin{equation}
S=\frac{1}{K}\sum_{k=1}^K P^k,
\end{equation}
where $K$ is the maximum length of the random walk. In what follows, we propose three approaches to define the hyperlink score based on $S$.

As a direct generalization of the superposed random walk index in Ref.~\cite{liu2010link}, the similarity between vertices $i$ and $j$ is given by $s_{ij}+s_{ji}$. Then the score of a hyperedge $e$ is the average of similarities between all pairs of vertices in the hyperedge; namely, the LRW index is defined as
\begin{equation}
S_e=\frac{2}{|e|(|e|-1)}\sum_{i < j, i\in e, j\in e}(s_{ij}+s_{ji}).\label{index1}
\end{equation}

A closer look at the matrix $S$ shows that it is right stochastic, that is, each row sums to one. In fact, the $i$-th row of $S$ corresponds to such a probability distribution: assuming that the walker currently at vertex $i$ first chooses the length of the walk $k$ uniformly random from $\{1,\cdots,K\}$ and then moves $k$ steps according to the transition matrix $P$, then $s_{ij}$ is the probability that the walker stops at vertex $j$. In general, for two similar vertices $i$ and $j$, the probabilities of random walkers starting from $i$ and $j$ reaching other vertices should be similar, which means that the probability distributions $S_{i}$ and $S_{j}$ should be close. As a consequence, we could measure similarity between $i$ and $j$ by characterizing the difference between the distributions $S_{i}$ and $S_{j}$. Here, we use the Jensen-Shannon divergence to measure it. Given the two probability distributions $p(x)$ and $q(x)$, the Jensen-Shannon divergence between them is defined as~\cite{lin1991divergence}
\begin{equation}
JS(p(x)||q(x)) = \frac{1}{2}\sum_x p(x)\log_2\frac{p(x)}{r(x)} + \frac{1}{2}\sum_x q(x)\log_2\frac{q(x)}{r(x)},
\end{equation}
where $r(x)=(p(x)+q(x))/2$. The Jensen-Shannon divergence is symmetric, i.e., $JS(p(x)||q(x))=JS(q(x)||p(x))$ and is bounded, i.e., $0\leqslant JS(p(x)||q(x))\leqslant 1$. Accordingly, the similarity between vertices $i$ and $j$ could be defined by $1-JS(S_{i}||S_{j})$ and the score of hyperedge $e$ is given by (the LRW-JS index)
\begin{equation}
S_e = 1 - \frac{2}{|e|(|e|-1)}\sum_{i < j, i\in e, j\in e}JS(S_{i}||S_{j}).\label{index2}
\end{equation}

Both the LRW and LRW-JS indirectly take the average similarity between all pairs of vertices as the hyperedge score. A possibility for scoring hyperedges in a direct way is to use the generalized Jensen-Shannon divergence, which allows direct comparison of more than two distributions. For $t$ distributions $p_1(x),p_2(x),\cdots,p_t(x)$ with weights $\pi=(\pi_1,\pi_2,\cdots,\pi_t)$ where $\pi_i\geqslant0$ and $\sum_i\pi_i=1$, the generalized Jensen-Shannon divergence is defined as~\cite{lin1991divergence}
\begin{equation}
JS_{\pi}(p_1(x),p_2(x),\cdots,p_t(x)) = \sum_{i}\pi_i\sum_x p_i(x)\log_2\frac{p_i(x)}{r(x)},
\end{equation}
where $r(x)=\sum_i\pi_ip_i(x)$. The generalized Jensen-Shannon divergence is invariant to any permutation of the distributions and is bounded above by $\log_2t$. Therefore, the third way we define the hyperlink score as (the LRW-GJS index)
\begin{equation}
S_e = 1-JS_{\pi_t}(S_{i_1},\cdots,S_{i_t})/\log_2t,\label{index3}
\end{equation}
where $t=|e|$, $\pi_t=(1/t,\cdots,1/t)$ and $e=\{i_1,\cdots,i_t\}$.

\section{Experiments}\label{experiment}

In this section, we evaluate the performance of the proposed indices on real-world datasets. Before presenting the results, we first briefly describe the datasets, the baseline methods and the evaluation metrics.

\begin{table}[h]\footnotesize
    \centering
    \caption{Basic properties of the datasets.}\label{data_info}
    \begin{tabular}{cccccc}
    \hline \hline
        & Datasets & No. vertices & No. hyperedges & Avg. vertex degree & Avg. edge size\\
        \hline
        (a) & Enron Email & 143 & 1457 & 31.43 & 3.09 \\
        (b) & Contact High School & 317 & 2320 & 22.70 & 3.10 \\
        (c) & NDC Classes & 628 & 794 & 9.02 & 7.14 \\
        (d) & iAF1260b & 1668 & 2047 & 5.27 & 4.30 \\
        (e) & DBLP & 4695 & 2346 & 2.85 & 5.69 \\
        (f) & Cora Co-reference & 1961 & 861 & 2.33 & 5.31 \\
        (g) & Cora Co-citation & 1330 & 1413 & 3.29 & 3.09 \\
    \hline
    \end{tabular}
\end{table}

\subsection{Datasets}

We consider $7$ datasets including the Enron Email, Contact High School, NDC Classes, iAF1260b, DBLP, Cora Co-reference and Co-citation. In data preprocessing, we remove duplicate hyperlinks and hyperlinks that contain less than two vertices. We only use the largest connected component of the resulted hypergraph in the experiments. A summary of the basic statistics of these datasets can be found in Table~\ref{data_info} and the corresponding description is presented below.

(a) Enron Email~\cite{klimt2004enron}. Each vertex is an email address and a hyperedge connects the sender's address and the receipt address.

(b) Contact High School~\cite{mastrandrea2015contact}. Each vertex represents a student in the high school and each hyperlink corresponds to a contact event, i.e., some students got close enough to each other. We only consider contacts involving at least three people.

(c) NDC Class~\cite{benson2018simplicial}. Vertices are class labels of drugs and each hyperedge is the set of class labels for a drug.

(d) iAF1260b~\cite{zhang2018beyond}. Each vertex represents a metabolite and each hyperedge connects the reactant and product metabolites in a chemical reaction.

(e) DBLP~\cite{ley2002dblp}. Each vertex is an author, and each hyperedge corresponds to a paper and connects its authors.

(f) Cora Co-reference and (g) Co-citation~\cite{sen2008collective}. In both datasets, vertices are papers. In the co-reference hypergraph, papers that refer to the same paper are connected by a hyperedge. In the co-citation hypergraph, a hyperedge connects papers cited by the same paper.

\subsection{Baselines}

We compare our indices with the following baselines, including three similarity indices, an optimization-based and a learning-based method.

(i) Hypergraph Common Neighbors (HCN) and Hypergraph Katz (HKatz). They are direct generalizations of the traditional Common Neighbors~\cite{newman2001clustering} and Katz~\cite{katz1953new}, respectively. Let $\mathcal{N}(i)$ be the set of neighbors of vertex $i$. As a local index, the HCN measures the similarity of $i$ and $j$ by the number of common neighbors, namely
\begin{equation}
    s_{ij}^{HCN}=|\mathcal{N}(i)\cap\mathcal{N}(j)|.
\end{equation}
Instead, the HKatz is a global index where the similarity between vertices is defined as
\begin{equation}
    s_{ij}^{HKatz}=\sum_{l=1}^\infty\beta^l(A^l)_{ij},
\end{equation}
where $A$ is the hypergraph adjacency matrix and $\beta$ is the damping factor. For the HCN and HKatz, the hyperedge score is given by the average of the similarities of all pairs of vertices in the hyperedge.

(ii) Hyperlink Prediction Using Resource Allocation with a Candidate Hyperedge Set (HPRA-CHS)~\cite{kumar2020hpra}. The HPRA-CHS considers resource allocation process in hypergraphs and defines the pair-wise similarity as
\begin{equation}
    s_{ij}^{HRA}=(W+WD_v^{-1}W)_{ij},
\end{equation}
where we remind that $W$ is the weighted adjacency matrix. Again, hyperedge scores are given by the average of the pair-wise scores.

(iii) Coordinated Matrix Minimization (CMM)~\cite{zhang2018beyond}. The CMM is an optimization-based method for hyperlink prediction. It tries to select some hyperlinks from the candidate set that are best suited to fill the training hypergraph by alternately performing matrix factorization and least square matching in adjacency space.\par

(iv) Hyperlink prediction via loop structure (HPLS)~\cite{pan2021predicting}. The HPLS exploits the hypergraph loop structure and adopts a modified logistic regression approach for hyperlink prediction.

\subsection{Setting and evaluating}

For each dataset, we randomly divide the hyperlinks into the observed hyperlink set $E^o$ and the missing hyperlink set $E^m$ by a ratio 4:1 (similar to Refs.~\cite{kumar2020hpra,liu2010link}) over 10 independent trials. The negative sampling strategy in Ref.~\cite{chen2022survey} is used to generate the set of fake hyperedges $E^{f}$: for each missing hyperedge $e$, we sample $\lambda=3$ corresponding fake hyperedges by randomly replacing $(1-\alpha)\cdot|e|$ vertices in $e$ with vertices in $V\backslash\{e\}$, similar to Refs.~\cite{zhang2018beyond,pan2021predicting,chen2022survey}. We also consider larger values of $\lambda$ and show the corresponding results in Appendix. Notice that the data splitting process may result in some vertices not being connected to any vertex in $E^{o}$, we exclude missing hyperlinks containing such vertices from $E^{m}$ and ensure that the sampled negative hyperlinks do not contain these vertices~\cite{kumar2020hpra}. The actual number of missing hyperlinks (on average) are 291, 462, 135, 293,
226, 85 and 173, respectively. Generally, the larger $\alpha$ is, the more difficult it is to identify the missing hyperlinks. In the experiments, we choose $\alpha$ from $\{0.2,0.5,0.8\}$. The candidate hyperlink set $E^{c}$ consists of the missing hyperlinks $E^{m}$ and the negative hyperlink set $E^{f}$. When scores of the candidate hyperlinks are computed, the $|E^{m}|$ hyperlinks with the highest scores are selected as predictions.

The parameter values are set as follows. For the HKatz, the damping factor $\beta$ is selected from $\{0.001,0.005,0.01,0.05,0.1\}$ using 5-fold cross-validation. For the CMM, we use the same values as in Ref.~\cite{zhang2018beyond}. The length cutoff $\tau_c$ of the HPLS is determined as follows: for datasets (a)-(c), $\tau_c$ is set to default 8 and for the remaining datasets, $\tau_c$ is selected from $\{6,7,\cdots,14\}$ using 2-fold cross validation. Our indices have one parameter: the maximum step $K$ of the random walks. We select $K$ from $\{2,3,4,5\}$ using 5-fold cross-validation. In $k$-fold cross-validation, the set of the observed hyperlinks $E^{o}$ is first divided into $k$ parts of equal size. Each time we select one of them as the missing hyperlink set and treat the others as observed hyperlinks. The process is repeated $k$ times so that each fold is used as the missing set once. The set of the candidate hyperlinks $E^{c}$ is used as the negative hyperlink set and the parameter with the highest average AUROC is selected.

We evaluate the performance by two popular metrics: area under the receiver operating characteristic curve (AUROC) and F1 score. The AUROC is widely used to evaluate the ability of a classification model to distinguish between classes. The F1 score is the harmonic mean of recall and precision, which reflects the prediction accuracy in a balanced way. Both the AUROC and F1 score range in value from 0 to 1, with higher values indicating better performance.

\begin{table}[h]\footnotesize
    \centering
    \caption{AUROC.}\label{res_auc}
    \begin{tabular}{cccccccccc}
    \hline\hline
        $\alpha$  & Dataset  & HCN & HKatz & HPRA-CHS & CMM & HPLS & LRW & LRW-JS & LRW-GJS \\
        \hline
        \multicolumn{1}{c}{\multirow{7}{*}{0.2}} & (a) & 0.8686 & 0.8810 & 0.9053 & 0.6314 & 0.8783 & 0.9153(5) & $\textbf{0.9301}$(2) & 0.9210(3)   \\
        \multicolumn{1}{c}{}                              & (b) & 0.9832 & 0.9819 & 0.9871 & 0.4330 & 0.9615 & 0.9917(5) & $\textbf{0.9934}$(2) & 0.9932(2)   \\
        \multicolumn{1}{c}{}                              & (c) & 0.9357 & 0.9019 & 0.9893 & 0.5329 & 0.8829 & $\textbf{0.9982}$(5) & 0.9966(2) & 0.9978(2)   \\
        \multicolumn{1}{c}{}                              & (d) & 0.9229 & 0.9224 & 0.9411 & 0.6639 & 0.9065 & $\textbf{0.9469}$(3) & 0.9017(2) & 0.9079(2)   \\
        \multicolumn{1}{c}{}                              & (e) & 0.9843 & 0.9803 & 0.9842 & 0.0998 & 0.9356 & 0.9856(5) & $\textbf{0.9878}$(5) & 0.9863(5)   \\
        \multicolumn{1}{c}{}                              & (f) & 0.7895 & 0.8383 & 0.8405 & 0.3889 & 0.6003 & 0.8914(5) & $\textbf{0.9048}$(5) & 0.8997(5)   \\
        \multicolumn{1}{c}{}                              & (g) & 0.8920 & 0.9118 & 0.9039 & 0.3715 & 0.8744 & $\textbf{0.9325}$(5) & 0.9267(2) & 0.9248(2)   \\ \hline
        \multicolumn{1}{c}{\multirow{7}{*}{0.5}} & (a) & 0.8190 & 0.8306 & 0.8546 & 0.6180 & 0.8378 & 0.8841(5) & $\textbf{0.9135}$(3) & 0.9036(4)   \\
        \multicolumn{1}{c}{}                              & (b) & 0.9718 & 0.9682 & 0.9746 & 0.4975 & 0.9644 & 0.9892(5) & $\textbf{0.9912}$(2) & 0.9902(2)   \\
        \multicolumn{1}{c}{}                              & (c) & 0.7957 & 0.7230 & 0.8417 & 0.5450 & 0.8492 & $\textbf{0.9887}$(4) & 0.9841(2) & 0.9835(2)   \\
        \multicolumn{1}{c}{}                              & (d) & 0.8007 & 0.7877 & 0.8215 & 0.5715 & 0.7829 & $\textbf{0.8773}$(2) & 0.8447(2) & 0.8511(2)   \\
        \multicolumn{1}{c}{}                              & (e) & 0.9141 & 0.9192 & 0.9272 & 0.2500 & 0.6844 & 0.9552(2) & $\textbf{0.9752}$(5) & 0.9716(5)   \\
        \multicolumn{1}{c}{}                              & (f) & 0.7673 & 0.8038 & 0.8251 & 0.4138 & 0.5695 & 0.8606(5) & $\textbf{0.8923}$(5) & 0.8861(5)   \\
        \multicolumn{1}{c}{}                              & (g) & 0.8475 & 0.8473 & 0.8522 & 0.4449 & 0.7115 & 0.8767(5) & $\textbf{0.8976}$(5) & 0.8945(5)   \\ \hline
        \multicolumn{1}{c}{\multirow{7}{*}{0.8}} & (a) & 0.7557 & 0.7397 & 0.7566 & 0.5729 & 0.7435 & 0.8005(4) & $\textbf{0.8733}$(5) & 0.8578(5)   \\
        \multicolumn{1}{c}{}                              & (b) & 0.8631 & 0.8017 & 0.8278 & 0.5058 & 0.8194 & 0.9096(5) & $\textbf{0.9419}$(3) & 0.9401(4)   \\
        \multicolumn{1}{c}{}                              & (c) & 0.6708 & 0.6303 & 0.6718 & 0.5082 & 0.7036 & 0.8930(2) & 0.9041(5) & $\textbf{0.9104}$(5)   \\
        \multicolumn{1}{c}{}                              & (d) & 0.6348 & 0.6214 & 0.6477 & 0.5325 & 0.6489 & $\textbf{0.7283}$(2) & 0.7254(2) & 0.7281(2)   \\
        \multicolumn{1}{c}{}                              & (e) & 0.7377 & 0.7459 & 0.7531 & 0.4106 & 0.4632 & 0.8087(2) & 0.8795(5) & $\textbf{0.8883}$(5)   \\
        \multicolumn{1}{c}{}                              & (f) & 0.6996 & 0.7347 & 0.7136 & 0.4306 & 0.5058 & 0.7457(5) & $\textbf{0.8116}$(5) & 0.8029(5)   \\
        \multicolumn{1}{c}{}                              & (g) & 0.7515 & 0.7368 & 0.7413 & 0.4823 & 0.5524 & 0.7672(5) & $\textbf{0.8325}$(5) & 0.8227(5)   \\
    \hline
    \end{tabular}
\end{table}

\subsection{Results and discussions}

\begin{table}[h]\footnotesize
    \centering
    \caption{F1 score.}\label{res_f1}
    \begin{tabular}{cccccccccc}
    \hline\hline
        $\alpha$  & Dataset  & HCN & HKatz & HPRA-CHS & CMM & HPLS & LRW & LRW-JS & LRW-GJS \\
        \hline
        \multicolumn{1}{c}{\multirow{7}{*}{0.2}} & (a) & 0.6629 & 0.7302 & 0.7312 & 0.2987 & 0.6981 & 0.7498 & $\textbf{0.7753}$ & 0.7698   \\
        \multicolumn{1}{c}{}                              & (b) & 0.8856 & 0.8785 & 0.8999 & 0.2701 & 0.8758 & 0.9270 & $\textbf{0.9435}$ & 0.9431   \\
        \multicolumn{1}{c}{}                              & (c) & 0.8047 & 0.6517 & 0.9023 & 0.2765 & 0.7344 & $\textbf{0.9841}$ & 0.9438 & 0.9659   \\
        \multicolumn{1}{c}{}                              & (d) & 0.8270 & 0.8299 & $\textbf{0.8740}$ & 0.4595 & 0.7799 & 0.8550 & 0.7644 & 0.7823   \\
        \multicolumn{1}{c}{}                              & (e) & 0.9487 & 0.9507 & 0.9619 & 0.0109 & 0.8093 & 0.9584 & $\textbf{0.9647}$ & 0.9620   \\
        \multicolumn{1}{c}{}                              & (f) & 0.6312 & 0.6716 & 0.7270 & 0.1265 & 0.3468 & 0.7767 & $\textbf{0.7949}$ & 0.7927   \\
        \multicolumn{1}{c}{}                              & (g) & 0.8120 & 0.8159 & 0.8202 & 0.1434 & 0.7316 & $\textbf{0.8371}$ & 0.8238 & 0.8168   \\ \hline
        \multicolumn{1}{c}{\multirow{7}{*}{0.5}} & (a) & 0.5884 & 0.6331 & 0.6499 & 0.2794 & 0.6577 & 0.7183 & $\textbf{0.7438}$ & 0.7348   \\
        \multicolumn{1}{c}{}                              & (b) & 0.8564 & 0.8312 & 0.8525 & 0.3359 & 0.8544 & 0.9168 & $\textbf{0.9316}$ & 0.9307   \\
        \multicolumn{1}{c}{}                              & (c) & 0.5701 & 0.4001 & 0.5730 & 0.3121 & 0.6420 & $\textbf{0.9122}$ & 0.8935 & 0.9092   \\
        \multicolumn{1}{c}{}                              & (d) & 0.5811 & 0.5298 & 0.5897 & 0.3029 & 0.6121 & $\textbf{0.7269}$ & 0.6870 & 0.7013   \\
        \multicolumn{1}{c}{}                              & (e) & 0.7506 & 0.7660 & 0.7732 & 0.1238 & 0.4661 & 0.8740 & $\textbf{0.9251}$ & 0.9226   \\
        \multicolumn{1}{c}{}                              & (f) & 0.5700 & 0.6122 & 0.6571 & 0.1516 & 0.3228 & 0.6715 & $\textbf{0.7708}$ & 0.7578   \\
        \multicolumn{1}{c}{}                              & (g) & 0.7175 & 0.6541 & 0.6968 & 0.1919 & 0.5182 & 0.7129 & $\textbf{0.7819}$ & 0.7746   \\ \hline
        \multicolumn{1}{c}{\multirow{7}{*}{0.8}} & (a) & 0.4604 & 0.4773 & 0.4728 & 0.2522 & 0.5336 & 0.5520 & $\textbf{0.6718}$ & 0.6477   \\
        \multicolumn{1}{c}{}                              & (b) & 0.6854 & 0.5551 & 0.6027 & 0.3239 & 0.6103 & 0.7307 & $\textbf{0.7781}$ & 0.7744   \\
        \multicolumn{1}{c}{}                              & (c) & 0.4286 & 0.2922 & 0.3484 & 0.2640 & 0.4448 & 0.6757 & 0.7799 & $\textbf{0.7867}$   \\
        \multicolumn{1}{c}{}                              & (d) & 0.3708 & 0.3349 & 0.4021 & 0.2394 & 0.4086 & 0.5128 & 0.5196 & $\textbf{0.5279}$   \\
        \multicolumn{1}{c}{}                              & (e) & 0.4780 & 0.4754 & 0.4833 & 0.1867 & 0.2331 & 0.5736 & 0.7144 & $\textbf{0.7326}$   \\
        \multicolumn{1}{c}{}                              & (f) & 0.4695 & 0.5026 & 0.4764 & 0.1452 & 0.2562 & 0.4817 & $\textbf{0.6131}$ & 0.6023   \\
        \multicolumn{1}{c}{}                              & (g) & 0.5390 & 0.4968 & 0.5076 & 0.2040 & 0.3185 & 0.5182 & $\textbf{0.6549}$ & 0.6348   \\
    \hline
    \end{tabular}
\end{table}

The AUROC and F1 score are shown in Tables~\ref{res_auc} and~\ref{res_f1} (the dataset labels (a)-(g) are the same as Table~\ref{data_info}), respectively. Each value in the tables is averaged over 10 independent trials. The numbers inside the brackets in Table~\ref{res_auc} are the most frequently selected $K$ values in cross-validation (By the way, we no longer report these $K$ values in Table~\ref{res_f1} since they are the same as those in Table~\ref{res_auc}). We have the following observations.

(i) All three of our proposed indices outperform state-of-the-art methods, with the LRW-JS achieving the best performance on most datasets and near-best performance on others. The LRW-GJS performs slightly worse than the LRW-JS but better than the LRW, especially when $\alpha$ is large. In generally, the LRW only uses partial information of the distributions when describing vertex similarity. In contrast, the LRW-JS and LRW-GJS characterize the vertex similarity and hyperedge score by the difference of corresponding probability distributions, utilizing the full information of these distributions. Therefore, we can expect the LRW-JS and LRW-GJS to perform better in hyperlink prediction.

(ii) The NDC Class and DBLP hypergraphs have relatively large average hyperedge size, which facilitates distinguishing missing hyperedges from fake hyperedges. Most baseline methods perform very well on these two datasets. Our proposed indices consistently beat the baselines in terms of both AUROC and F1 score by a narrow margin.

(iii) In the iAF1260b dataset, the LRW has the best performance. In contrast, the LRW-JS and LRW-GJS are less effective, and even perform worse than some baselines in cade of $\alpha=0.2$. The main reason is that in this hypernetwork several vertices have very large hyperdegrees (the top five vertex hyperdegrees are 879, 521, 334, 266, and 262, respectively), while most vertices have very low hyperdegrees (1053 vertices have hyperdegrees no more than 2), which makes many vertices very similar to each other when measured with the (generalized) Jensen-Shannon divergence. In this case, the LRW is more applicable than the LRW-JS and LRW-GJS.

(iv) The remaining four datasets include two relatively dense networks, Enron Email and Contact High School, and two sparse networks, Cora Co-reference and Co-citation. The superiority of our indices over baseline methods in these datasets indicates that they are effective for both sparse and dense hypergraphs.

(v) The parameter $\alpha$ controls the separability of missing hyperlinks and negative hyperlinks. As $\alpha$ becomes larger, which means that predicting missing hyperlinks becomes more difficult, our proposed indices still outperform the baselines, and the advantage is more pronounced. Moreover, the superiority of the LRW-JS and LRW-GJS over the LRW also demonstrates the effectiveness of the (generalized) Jensen-Shannon divergence for measuring vertex similarity and hyperedge score.

\begin{figure}[h]
\centering
    \begin{minipage}[t]{0.475\textwidth}
    \centering
    \includegraphics[width=\textwidth]{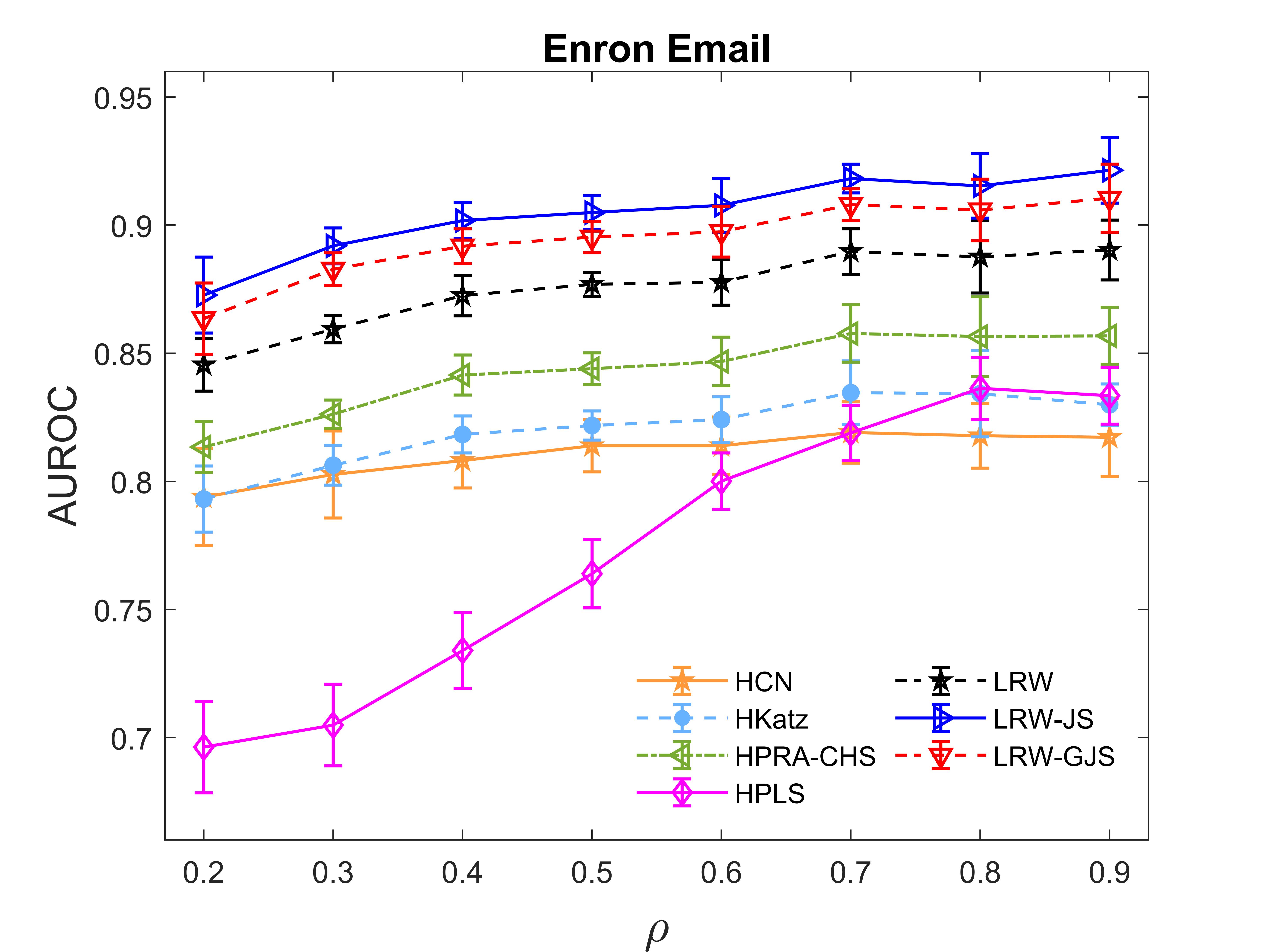}
    \end{minipage}
    \hfill
    \begin{minipage}[t]{0.475\textwidth}
    \centering
    \includegraphics[width=\textwidth]{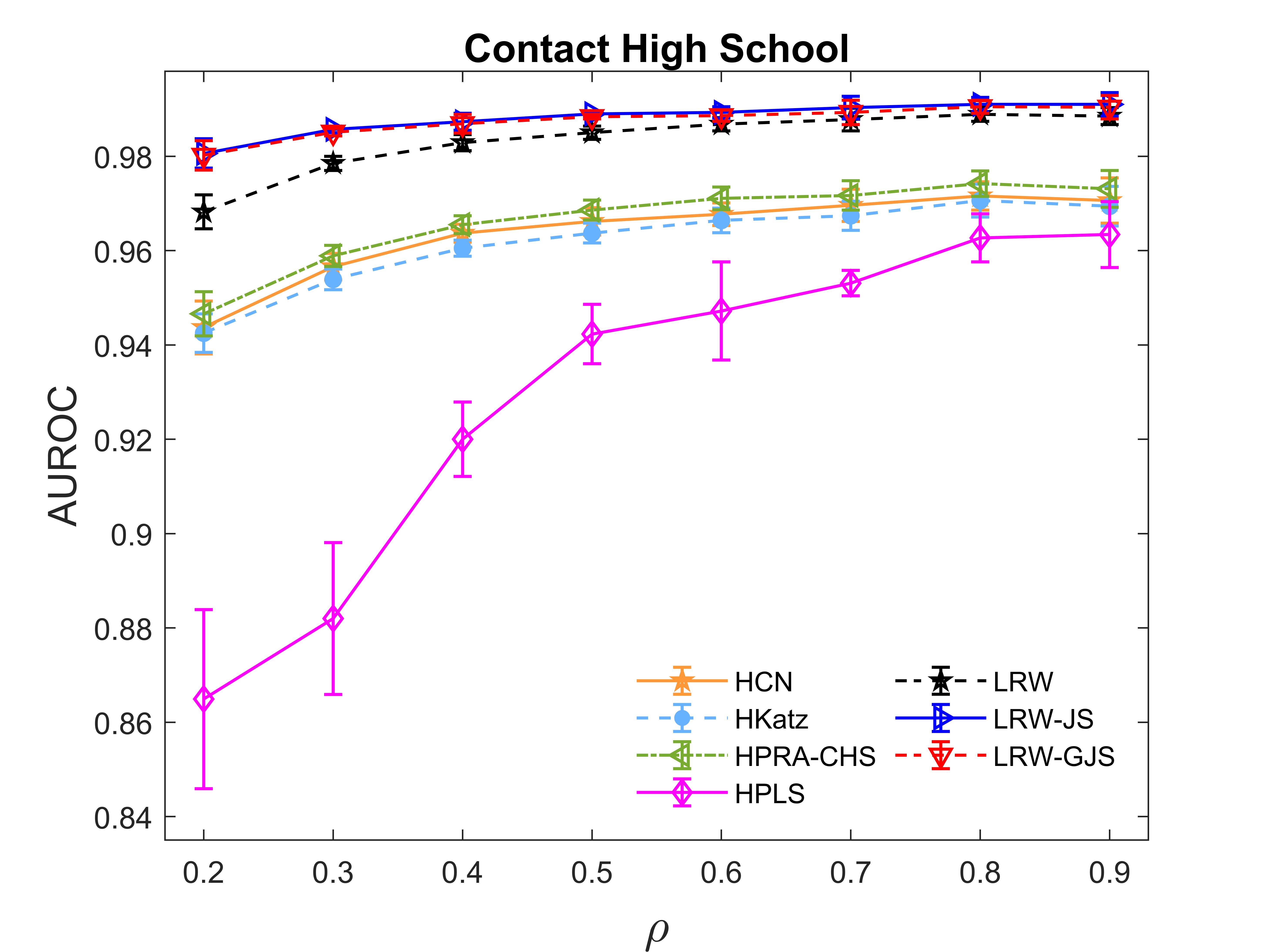}
    \end{minipage}
    \vskip\baselineskip
    \begin{minipage}[t]{0.475\textwidth}
    \centering
    \includegraphics[width=\textwidth]{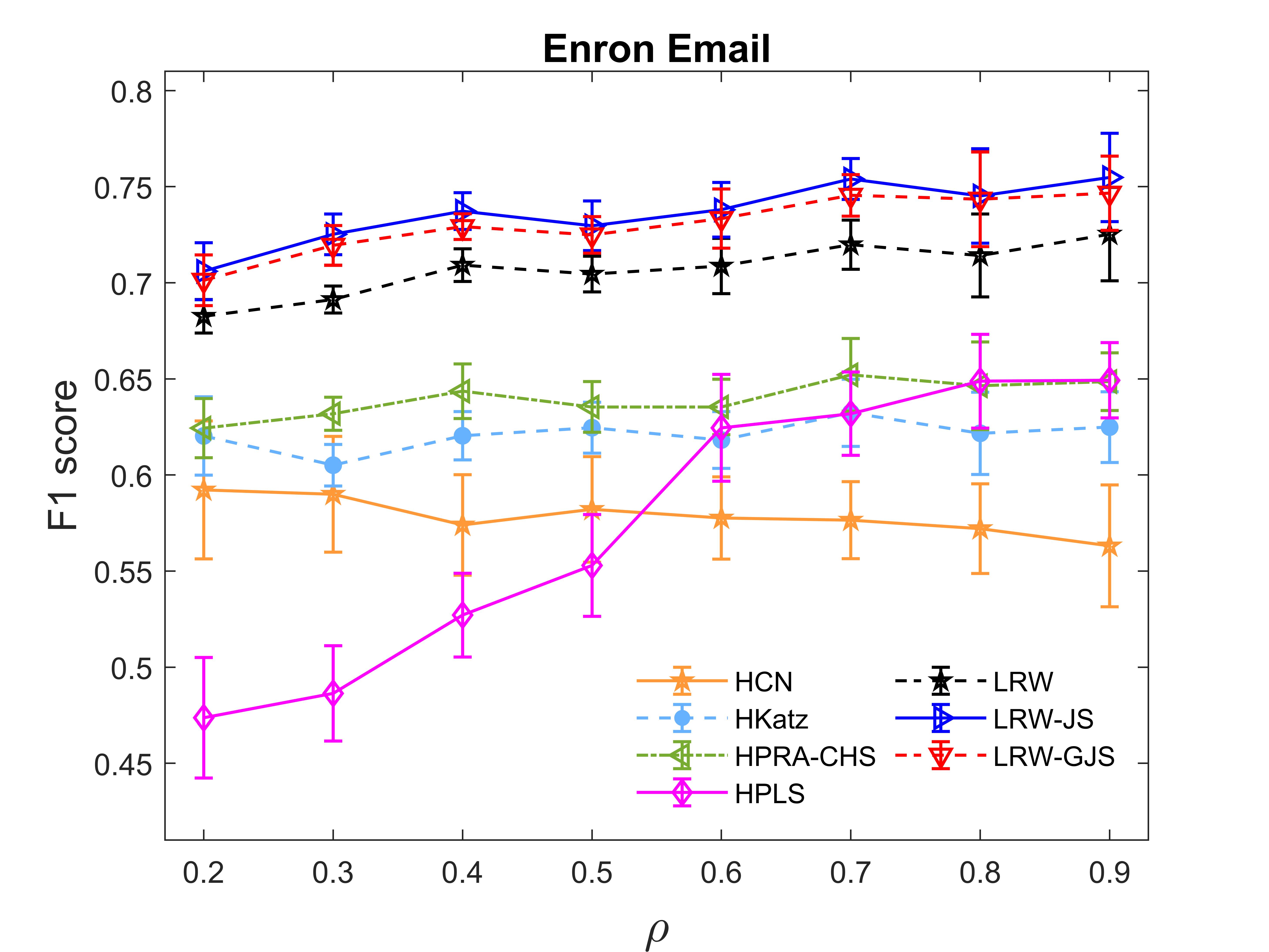}
    \end{minipage}
    \hfill
    \begin{minipage}[t]{0.475\textwidth}
    \centering
    \includegraphics[width=\textwidth]{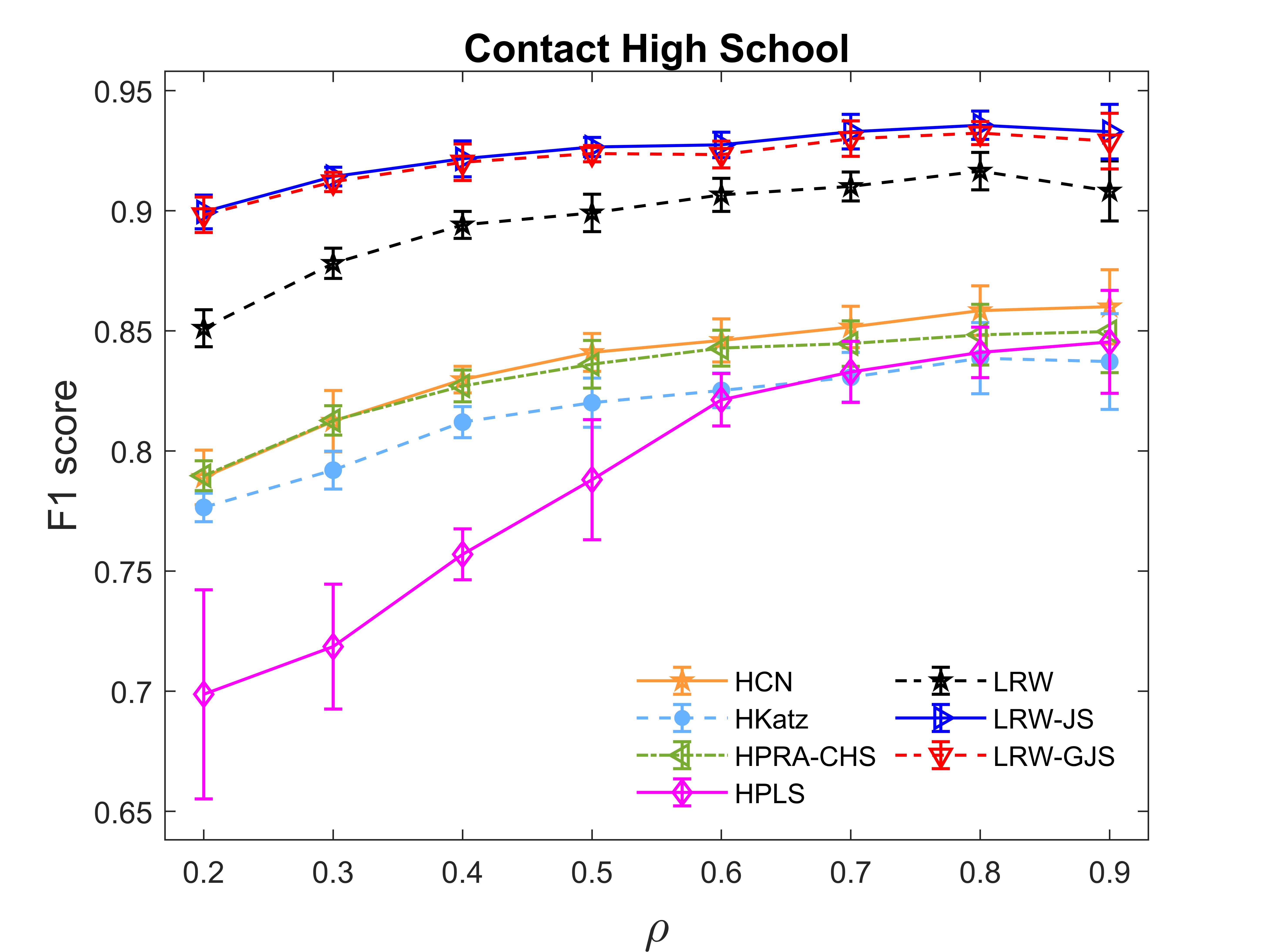}
    \end{minipage}
\caption{(Color online) Dependence of the AUROC and F1 score on the fraction of observed hyperlinks in Enron Email and Contact High School.}
\label{fig1}
\end{figure}

Finally, we investigate the impact the fraction of observed hyperlinks $\rho$ on prediction. We only conduct experiments on the two relatively dense hypergraphs, Enron Email and Contact High School. For other datasets, the observed hypergraph will consist of many disconnected components if $\rho$ is too small. We fix $\alpha$ to be $0.5$ and vary $\rho$ from $0.2$ to $0.9$. For each dataset, we run 10 independent trials. As shown in Fig.~\ref{fig1}, our proposed indices consistently perform the best under different $\rho$ values. The results of the CMM are not present because its performance is much worse than the others.

The time complexity of the calculation of $S$ is approximately $O(nd^K)$ where $d$ is the average degree of the clique expansion of the hypergraph. For sparse hypergraphs with small hyperedge cardinality, the clique expansion graph is usually sparse and the calculation of $S$ is very efficient. For the LRW, the calculation of the score of hyperedge $e$ need $O(|e|^2)$ time. Therefore the total time complexity of the LRW is $O(nd^K+qs^2)$ where $q=|E^{c}|$ is the number of candidate hyperlinks and $s$ is the average size of the candidate hyperedges. The Jensen-Shannon divergence can be calculated in a time $O(d^K)$ since there are $O(d^K)$ non-zero elements in the rows of $S$. For the LRW-JS, we have two ways to calculate the hyperedge score. In the first approach, we pre-compute all pair-wise similarity and return the score of hyperedge $e$ in $O(|e|^2)$ time. In the second approach, we perform $O(|e|^2)$ calculations of the Jensen-Shannon divergence for each candidate hyperedge $e$. Therefore the computational complexity of the LRW-JS is $O(nd^K+\min(n^2d^K+qs^2,qs^2d^K))$. As for the LRW-GJS, we only adopt the second approach, hence its time complexity is $O((n+qs^2)d^K)$ where the calculation of the generalized Jensen-Shannon entropy for $s$ vertices requires $O(s^2d^K)$ time. We conclude that among the proposed three indices, the LRW is faster but less effective, whereas the LRW-JS and LRW-GJS are usually more effective at the cost of higher computational complexity.

\section{Conclusion}\label{conclusion}

In this paper, we have considered the novel problem of hyperlink prediction. The basic idea is to endue each vertex with a probability distribution and compare the differences between these distributions to characterize vertex similarities and hyperedge scores. For this purpose, we first employ random walks on hypergraphs to obtain a probability distribution $S_{i}$ for vertex $i$ which reflects how likely a random walker at $i$ is to reach each vertex in local moves. Then, we introduce three indices to depict hyperedge scores. For the LRW and LRW-JS, we first measure pair similarity of vertices and then score a hyperedge by the average of the similarities of all pairs of vertices in the hyperedge. When measuring similarity between $i$ and $j$, the LRW trivially makes use of the transition probability $s_{ij} + s_{ji}$ (see Eq.~(\ref{index1})), which only considers partial information of distributions $S_i$ and $S_j$. Instead, the LRW-JS characterizes the similarity as $1-JS(S_i,S_j)$ (see Eq.~(\ref{index2})), which considers full information contained in these distributions. Different from these two indirct indices, the LRW-GJS, directly scores the hyperedge by the difference (generalized JS divergence) of distributions corresponding to vertices in the hyperedge (see Eq.~(\ref{index3})). 

Among the baseline methods, the HKatz, CMM and HPLS have high computational complexity which greatly limits their applicability. Moreover, the baselines usually have low prediction accuracy. Our indices have the following advantages: (i) The proposed indices have high accuracy and low computational complexity, especially compared to non-similarity-based methods \cite{xu2013hyperlink,zhang2018beyond,sharma2021c3mm,pan2021predicting}; (ii) The probability distributions are topological features \cite{pan2021predicting}, which are more interpretable than latent features \cite{zhang2018beyond}; and (iii) By taking advantage of the (generalized) JS divergence, both the LRW-JS and LRW-GJS benefit from information theory. As a result, our indices show their superiority in a wide range of datasets.

We suggest three future extensions. First, there are several definitions of random walks on hypergraphs~\cite{carletti2020random,chitra2019random} and it is interesting to compare them under the hyperlink prediction task, which can give us more insight into their differences and applicability. Second, the LRW-JS and LRW-GJS perform well for most datasets except for iAF1260b, which is mainly due to the presence of a few hubs and many low-degree vertices. The proposed indices deserve further improvement. Finally, it has been shown that information entropy can better capture the topological difference than the other typical network measurements~\cite{bianconi2009}. Therefore, it is natural to consider that the probability of a missing link between two vertices can be transformed into the corresponding information entropy.

The application scenarios of the proposed hyperlink prediction methods include missing reaction inference in metabolic networks~\cite{zhang2018beyond}, potential friend recommendation in online social networks~\cite{aiello2012friendship}, spam mail detection in email networks~\cite{huang2006link}, etc.

\section*{Acknowledgments}

This work was supported by the Natural Science Foundation of China under Grant No. 12071281 and Science and Technology Commission of Shanghai Municipality under Grant No. 22JC1401401.

\section*{Appendix}\label{append}
Here follows the AUROC and F1 score in cases of $\lambda=10$ and $100$, respectively. The missing entries correspond to experiments that run out of memory, or that have not ended after $24$ hours of execution. Our proposed indices still perform better than baseline methods even when fake hyperedges are more.

\begin{table}[h]\footnotesize
    \centering
    \caption{AUROC ($\lambda=10$).}\label{res_auc_b10}
    \begin{tabular}{cccccccccc}
    \hline\hline
        $\alpha$  & Dataset  & HCN & HKatz & HPRA-CHS & CMM & HPLS & LRW & LRW-JS & LRW-GJS \\
        \hline
        \multicolumn{1}{c}{\multirow{7}{*}{0.2}} & (a) & 0.8738 & 0.8853 & 0.9082 & 0.6423 & 0.8900 & 0.9189(5) & $\textbf{0.9319}$(2) & 0.9241(3)   \\
        \multicolumn{1}{c}{}                              & (b) & 0.9816 & 0.9842 & 0.9858 & 0.4116 & 0.9713 & 0.9915(5) & $\textbf{0.9935}$(2) & 0.9933(2)   \\
        \multicolumn{1}{c}{}                              & (c) & 0.9349 & 0.9012 & 0.9897 & 0.4807 & 0.9071 & $\textbf{0.9979}$(2) & 0.9966(2) & 0.9977(2)   \\
        \multicolumn{1}{c}{}                              & (d) & 0.9316 & 0.9321 & 0.9401 & 0.6451 & 0.9092 & $\textbf{0.9457}$(3) & 0.9052(2) & 0.9115(2)   \\
        \multicolumn{1}{c}{}                              & (e) & 0.9793 & 0.9771 & 0.9808 & - & 0.9539 & 0.9805(5) & $\textbf{0.9828}$(3) & 0.9804(4)   \\
        \multicolumn{1}{c}{}                              & (f) & 0.7996 & 0.8417 & 0.8560 & 0.4101 & 0.6630 & 0.9075(5) & $\textbf{0.9165}$(5) & 0.9127(5)   \\
        \multicolumn{1}{c}{}                              & (g) & 0.9054 & 0.9227 & 0.9096 & 0.3703 & 0.8920 & $\textbf{0.9325}$(5) & 0.9296(2) & 0.9275(2)   \\ \hline
        \multicolumn{1}{c}{\multirow{7}{*}{0.5}} & (a) & 0.8181 & 0.8323 & 0.8562 & 0.6123 & 0.8472 & 0.8867(4) & $\textbf{0.9152}$(3) & 0.9052(4)   \\
        \multicolumn{1}{c}{}                              & (b) & 0.9718 & 0.9715 & 0.9750 & 0.5068 & 0.9641 & 0.9894(5) & $\textbf{0.9914}$(2) & 0.9909(2)   \\
        \multicolumn{1}{c}{}                              & (c) & 0.7967 & 0.7273 & 0.8370 & 0.5344 & 0.8714 & $\textbf{0.9901}$(2) & 0.9879(2) & 0.9861(2)   \\
        \multicolumn{1}{c}{}                              & (d) & 0.8108 & 0.7963 & 0.8230 & 0.4890 & 0.8176 & $\textbf{0.8770}$(2) & 0.8469(2) & 0.8531(2)   \\
        \multicolumn{1}{c}{}                              & (e) & 0.9116 & 0.9132 & 0.9265 & - & 0.8565 & 0.9533(2) & $\textbf{0.9671}$(5) & 0.9644(5)   \\
        \multicolumn{1}{c}{}                              & (f) & 0.7724 & 0.8231 & 0.8204 & 0.4264 & 0.5860 & 0.8666(5) & $\textbf{0.9063}$(5) & 0.8989(5)   \\
        \multicolumn{1}{c}{}                              & (g) & 0.8635 & 0.8617 & 0.8611 & 0.4770 & 0.8014 & 0.8823(5) & $\textbf{0.9074}$(5) & 0.9036(5)   \\ \hline
        \multicolumn{1}{c}{\multirow{7}{*}{0.8}} & (a) & 0.7436 & 0.7328 & 0.7486 & 0.5541 & 0.7531 & 0.7915(4) & $\textbf{0.8690}$(4) & 0.8532(5)   \\
        \multicolumn{1}{c}{}                              & (b) & 0.8652 & 0.8130 & 0.8298 & 0.5426 & 0.8233 & 0.9088(5) & $\textbf{0.9399}$(3) & 0.9384(4)   \\
        \multicolumn{1}{c}{}                              & (c) & 0.6692 & 0.6383 & 0.6768 & 0.5203 & 0.6989 & 0.8888(2) & 0.9049(5) & $\textbf{0.9111}$(5)   \\
        \multicolumn{1}{c}{}                              & (d) & 0.6397 & 0.6262 & 0.6496 & 0.5582 & 0.6490 & 0.7256(2) & 0.7227(2) & $\textbf{0.7264}$(2)   \\
        \multicolumn{1}{c}{}                              & (e) & 0.7339 & 0.7444 & 0.7571 & - & 0.5715 & 0.8183(2) & 0.8814(5) & $\textbf{0.8890}$(5)   \\
        \multicolumn{1}{c}{}                              & (f) & 0.6878 & 0.7372 & 0.7274 & 0.4508 & 0.4856 & 0.7613(5) & $\textbf{0.8347}$(5) & 0.8258(5)   \\
        \multicolumn{1}{c}{}                              & (g) & 0.7589 & 0.7434 & 0.7438 & 0.4840 & 0.5881 & 0.7675(5) & $\textbf{0.8336}$(5) & 0.8245(5)   \\
    \hline
    \end{tabular}
\end{table}

\begin{table}[h]\footnotesize
    \centering
    \caption{F1 score ($\lambda=10$).}\label{res_f1_b10}
    \begin{tabular}{cccccccccc}
    \hline\hline
        $\alpha$  & Dataset  & HCN & HKatz & HPRA-CHS & CMM & HPLS & LRW & LRW-JS & LRW-GJS \\
        \hline
        \multicolumn{1}{c}{\multirow{7}{*}{0.2}} & (a) & 0.4615 & 0.5901 & 0.5990 & 0.2302 & 0.5949 & 0.5891 & 0.5953 & $\textbf{0.6087}$   \\
        \multicolumn{1}{c}{}                              & (b) & 0.8335 & 0.8031 & 0.8166 & 0.1711 & 0.8099 & 0.8691 & 0.8983 & $\textbf{0.8988}$   \\
        \multicolumn{1}{c}{}                              & (c) & 0.7106 & 0.4856 & 0.8407 & 0.2429 & 0.7160 & $\textbf{0.9453}$ & 0.9003 & 0.9321   \\
        \multicolumn{1}{c}{}                              & (d) & 0.7621 & 0.8062 & $\textbf{0.8438}$ & 0.4907 & 0.7553 & 0.7915 & 0.6852 & 0.7211   \\
        \multicolumn{1}{c}{}                              & (e) & 0.9080 & 0.9388 & 0.9484 & - & 0.8605 & 0.9437 & $\textbf{0.9512}$ & 0.9482   \\
        \multicolumn{1}{c}{}                              & (f) & 0.3692 & 0.5879 & 0.6810 & 0.0460 & 0.2745 & $\textbf{0.7201}$ & 0.6804 &0.7053   \\
        \multicolumn{1}{c}{}                              & (g) & 0.7452 & 0.7478 & 0.7584 & 0.0757 & 0.6971 & $\textbf{0.7647}$ & 0.7319 & 0.7432   \\ \hline
        \multicolumn{1}{c}{\multirow{7}{*}{0.5}} & (a) & 0.3625 & 0.4756 & 0.5065 & 0.2268 & 0.5268 & 0.5529 & 0.5584 & $\textbf{0.5732}$   \\
        \multicolumn{1}{c}{}                              & (b) & 0.7921 & 0.7416 & 0.7616 & 0.1961 & 0.7559 & 0.8604 & $\textbf{0.8858}$ & 0.8814   \\
        \multicolumn{1}{c}{}                              & (c) & 0.4973 & 0.2803 & 0.3781 & 0.2230 & 0.5887 & 0.8244 & 0.8740 & $\textbf{0.8858}$   \\
        \multicolumn{1}{c}{}                              & (d) & 0.4261 & 0.3667 & 0.4438 & 0.3029 & 0.1775 & $\textbf{0.6212}$ & 0.5672 & 0.6085   \\
        \multicolumn{1}{c}{}                              & (e) & 0.6261 & 0.5761 & 0.6302 & - & 0.5963 & 0.8085 & $\textbf{0.8914}$ & 0.8913   \\
        \multicolumn{1}{c}{}                              & (f) & 0.3152 & 0.5462 & 0.5446 & 0.0632 & 0.1769 & 0.5715 & 0.6631 & $\textbf{0.6862}$   \\
        \multicolumn{1}{c}{}                              & (g) & 0.6033 & 0.5655 & 0.5784 & 0.1627 & 0.5082 & 0.6216 & 0.6785 & $\textbf{0.6919}$   \\ \hline
        \multicolumn{1}{c}{\multirow{7}{*}{0.8}} & (a) & 0.2768 & 0.2920 & 0.3040 & 0.1821 & 0.3734 & 0.3882 & 0.4659 & $\textbf{0.4684}$   \\
        \multicolumn{1}{c}{}                              & (b) & 0.5391 & 0.3820 & 0.4268 & 0.1533 & 0.4383 & 0.5881 & $\textbf{0.5948}$ & 0.5894   \\
        \multicolumn{1}{c}{}                              & (c) & 0.3142 & 0.2177 & 0.2255 & 0.2222 & 0.3208 & 0.4748 & $\textbf{0.7319}$ &0.7266   \\
        \multicolumn{1}{c}{}                              & (d) & 0.2002 & 0.1859 & 0.1984 & 0.1419 & 0.2357 & 0.3260 & 0.3494 & $\textbf{0.3840}$   \\
        \multicolumn{1}{c}{}                              & (e) & 0.2874 & 0.2511 & 0.2771 & - & 0.1548 & 0.4367 & $\textbf{0.6569}$ & 0.6520   \\
        \multicolumn{1}{c}{}                              & (f) & 0.1955 & 0.3890 & 0.3352 & 0.1111 & 0.1051 & 0.3551 & $\textbf{0.5413}$ & 0.5250   \\
        \multicolumn{1}{c}{}                              & (g) & 0.3619 & 0.3129 & 0.3237 & 0.1429 & 0.1873 & 0.3739 & $\textbf{0.5421}$ & 0.5249   \\
    \hline
    \end{tabular}
\end{table}

\begin{table}[h]\footnotesize
    \centering
    \caption{AUROC ($\lambda=100$).}\label{res_auc_b100}
    \begin{tabular}{cccccccccc}
    \hline\hline
        $\alpha$  & Dataset  & HCN & HKatz & HPRA-CHS & CMM & HPLS & LRW & LRW-JS & LRW-GJS \\
        \hline
        \multicolumn{1}{c}{\multirow{7}{*}{0.2}} & (a) & 0.8679 & 0.8797 & 0.9051 & - & 0.8377 & 0.9172(5) & $\textbf{0.9323}$(2) & 0.9247(3)   \\
        \multicolumn{1}{c}{}                              & (b) & 0.9809 & 0.9841 & 0.9858 & - & 0.9687 & 0.9912(5) & $\textbf{0.9919}$(2) & 0.9918(2)   \\
        \multicolumn{1}{c}{}                              & (c) & 0.9288 & 0.8975 & 0.9864 & - & 0.8990 & $\textbf{0.9970}$(2) & 0.9942(2) & 0.9959(2)   \\
        \multicolumn{1}{c}{}                              & (d) & 0.9272 & 0.9316 & 0.9434 & - & 0.9059 & $\textbf{0.9476}$(3) & 0.8964(2) & 0.9035(2)   \\
        \multicolumn{1}{c}{}                              & (e) & 0.9802 & 0.9819 & 0.9819 & - & - & 0.9868(5) & $\textbf{0.9898}$(4) & 0.9880(4)   \\
        \multicolumn{1}{c}{}                              & (f) & 0.8028 & 0.8387 & 0.8544 & - & 0.6762 & 0.8900(5) & $\textbf{0.8984}$(5) & 0.8967(5)   \\
        \multicolumn{1}{c}{}                              & (g) & 0.9012 & 0.9262 & 0.9049 & - & 0.8887 & $\textbf{0.9393}$(5) & 0.9345(2) & 0.9315(2)   \\ \hline
        \multicolumn{1}{c}{\multirow{7}{*}{0.5}} & (a) & 0.8137 & 0.8262 & 0.8509 & - & 0.8133 & 0.8841(4) & $\textbf{0.9183}$(3) & 0.9081(4)   \\
        \multicolumn{1}{c}{}                              & (b) & 0.9709 & 0.9710 & 0.9743 & - & 0.9609 & 0.9886(5) & $\textbf{0.9898}$(2) & 0.9894(2)   \\
        \multicolumn{1}{c}{}                              & (c) & 0.7941 & 0.7251 & 0.8311 & - & 0.8366 & $\textbf{0.9894}$(2) & 0.9853(2) & 0.9851(2)   \\
        \multicolumn{1}{c}{}                              & (d) & 0.8079 & 0.8004 & 0.8281 & - & 0.7818 & $\textbf{0.8737}$(2) & 0.8422(2) & 0.8489(2)   \\
        \multicolumn{1}{c}{}                              & (e) & 0.9123 & 0.9192 & 0.9274 & - & - & 0.9553(2) & $\textbf{0.9768}$(4) & 0.9724(5)   \\
        \multicolumn{1}{c}{}                              & (f) & 0.7764 & 0.8098 & 0.8223 & - & 0.5858 & 0.8597(5) & $\textbf{0.8928}$(5) & 0.8881(5)   \\
        \multicolumn{1}{c}{}                              & (g) & 0.8572 & 0.8620 & 0.8560 & - & 0.8097 & 0.8862(5) & $\textbf{0.9132}$(5) & 0.9091(5)   \\ \hline
        \multicolumn{1}{c}{\multirow{7}{*}{0.8}} & (a) & 0.7404 & 0.7312 & 0.7483 & - & 0.6753 & 0.7917(4) & $\textbf{0.8737}$(4) & 0.8578(5)   \\
        \multicolumn{1}{c}{}                              & (b) & 0.8613 & 0.8097 & 0.8263 & - & 0.8154 & 0.9059(5) & $\textbf{0.9357}$(3) & 0.9338(3)   \\
        \multicolumn{1}{c}{}                              & (c) & 0.6665 & 0.6354 & 0.6710 & - & 0.7237 & 0.8893(2) & 0.9023(4) & $\textbf{0.9092}$(4)   \\
        \multicolumn{1}{c}{}                              & (d) & 0.6328 & 0.6249 & 0.6465 & - & 0.6299 & 0.7156(2) & 0.7177(2) & $\textbf{0.7204}$(2)   \\
        \multicolumn{1}{c}{}                              & (e) & 0.7310 & 0.7443 & 0.7562 & - & - & 0.8121(2) & 0.8864(5) & $\textbf{0.8954}$(5)   \\
        \multicolumn{1}{c}{}                              & (f) & 0.6952 & 0.7401 & 0.7258 & - & 0.4552 & 0.7560(5) & $\textbf{0.8310}$(5) & 0.8242(5)   \\
        \multicolumn{1}{c}{}                              & (g) & 0.7502 & 0.7391 & 0.7346 & - & 0.6410 & 0.7644(5) & $\textbf{0.8383}$(5) & 0.8282(5)   \\
    \hline
    \end{tabular}
\end{table}

\begin{table}[h]\footnotesize
    \centering
    \caption{F1 score ($\lambda=100$).}\label{res_f1_b100}
    \begin{tabular}{cccccccccc}
    \hline\hline
        $\alpha$  & Dataset  & HCN & HKatz & HPRA-CHS & CMM & HPLS & LRW & LRW-JS & LRW-GJS \\
        \hline
        \multicolumn{1}{c}{\multirow{7}{*}{0.2}} & (a) & 0.1615 & 0.2440 & 0.2570 & - & $\textbf{0.3526}$ & 0.1677 & 0.1512 & 0.1766   \\
        \multicolumn{1}{c}{}                              & (b) & 0.6331 & 0.5390 & 0.5743 & - & 0.6220 & 0.6532 & 0.6771 & $\textbf{0.6773}$   \\
        \multicolumn{1}{c}{}                              & (c) & 0.5144 & 0.4155 & 0.6478 & - & 0.6713 & $\textbf{0.7409}$ & 0.6739 & 0.7312   \\
        \multicolumn{1}{c}{}                              & (d) & 0.6033 & 0.6100 & 0.6667 & - & $\textbf{0.6733}$ & 0.6000 & 0.4633 & 0.5400   \\
        \multicolumn{1}{c}{}                              & (e) & 0.8226 & 0.8795 & 0.8974 & - & - & 0.8815 & 0.8816 & $\textbf{0.9090}$   \\
        \multicolumn{1}{c}{}                              & (f) & 0.0641 & 0.2255 & 0.4896 & - & 0.1492 & $\textbf{0.5162}$ & 0.3846 & 0.4092   \\
        \multicolumn{1}{c}{}                              & (g) & 0.5342 & 0.5761 & $\textbf{0.5937}$ & - & 0.5034 & 0.5512 & 0.4536 & 0.4822   \\ \hline
        \multicolumn{1}{c}{\multirow{7}{*}{0.5}} & (a) & 0.0887 & 0.1821 & 0.1959 & - & $\textbf{0.2962}$ & 0.1821 & 0.1526 & 0.1828   \\
        \multicolumn{1}{c}{}                              & (b) & 0.5534 & 0.4567 & 0.5098 & - & 0.5680 & $\textbf{0.6536}$ & 0.6431 & 0.6449   \\
        \multicolumn{1}{c}{}                              & (c) & 0.3939 & 0.2269 & 0.2295 & - & 0.5022 & 0.4955 & 0.7130 & $\textbf{0.7643}$   \\
        \multicolumn{1}{c}{}                              & (d) & 0.2233 & 0.1702 & 0.2201 & - & 0.2400 & 0.3633 & 0.2933 & $\textbf{0.3822}$   \\
        \multicolumn{1}{c}{}                              & (e) & 0.4390 & 0.3465 & 0.3789 & - & - & 0.5896 & 0.8145 & $\textbf{0.8574}$   \\
        \multicolumn{1}{c}{}                              & (f) & 0.0666 & 0.3309 & 0.3781 & - & 0.0817 & 0.4189 & 0.4084 & $\textbf{0.4276}$   \\
        \multicolumn{1}{c}{}                              & (g) & 0.2541 & 0.2384 & 0.2435 & - & 0.2988 & 0.3958 & 0.3890 & $\textbf{0.4098}$   \\ \hline
        \multicolumn{1}{c}{\multirow{7}{*}{0.8}} & (a) & 0.0591 & 0.0989 & 0.1109 & - & 0.0976 & 0.1168 & 0.1093 & $\textbf{0.1182}$   \\
        \multicolumn{1}{c}{}                              & (b) & 0.1382 & 0.1102 & 0.1598 & - & 0.1944 & $\textbf{0.2336}$ & 0.2126 & 0.2078   \\
        \multicolumn{1}{c}{}                              & (c) & 0.1560 & 0.1176 & 0.1028 & - & 0.1107 & 0.2686 & 0.5391 & $\textbf{0.5509}$   \\
        \multicolumn{1}{c}{}                              & (d) & 0.0867 & 0.0733 & 0.0767 & - & 0.0867 & 0.0900 & 0.0767 & $\textbf{0.1067}$   \\
        \multicolumn{1}{c}{}                              & (e) & 0.1097 & 0.0787 & 0.0854 & - & - & 0.1461 & 0.5193 & $\textbf{0.5315}$   \\
        \multicolumn{1}{c}{}                              & (f) & 0.0461 & 0.1447 & 0.1231 & - & 0.0162 & 0.1253 & $\textbf{0.3473}$ & 0.3452   \\
        \multicolumn{1}{c}{}                              & (g) & 0.0791 & 0.0670 & 0.0891 & - & 0.0760 & 0.1434 & $\textbf{0.2971}$ & 0.2959   \\
    \hline
    \end{tabular}
\end{table}

\end{document}